# Public Acceptance of Various Ways to Use and Treat AI: A Predictive Moral Framework

**Authors:** Kimmo Eriksson[1,2*], Simon Karlsson[1], Irina Vartanova[1,3], Pontus Strimling[1,4]

[1]Institute for Futures Studies, Stockholm, Sweden
[2]School of Education, Culture and Communication, Mälardalen University, Västerås, Sweden
[3]Department of Women's and Children's Health, Uppsala University
[4]Institute for Analytical Sociology, Linköping University, Norrköping, Sweden

*Corresponding author: Kimmo Eriksson, Mälardalen University, Box 883, SE-72123 Stockholm, Sweden. E-mail: kimmo.eriksson@mdu.se

***Funding***

This research was funded by the Knut and Alice Wallenberg Foundation (grant no. 2022.0191, recipient PS) and a Mälardalen University AI and Society Research Fellowship (recipient KE).

***Author Contributions***

KE and PS conceived the research. SK collected the data. SK and IV performed the analyses with input from KE. KE wrote the paper. All authors reviewed and approved the paper.

**Abstract**

As artificial intelligence rapidly transforms society, developers and policymakers struggle to anticipate which applications will face public moral resistance. We propose that the public's judgments are predictable. In a large, preregistered study (N = 589, U.S. representative sample), we used a comprehensive taxonomy of 100 AI applications spanning personal and organizational contexts—including both functional uses and the moral treatment of AI itself. In participants' collective judgment, applications ranged from highly unacceptable to fully acceptable. This variation was very accurately predicted using measures of applications' possession of five relatively independent moral qualities: perceived risk, benefit, dishonesty, unnaturalness, and reduced accountability. This framework also predicted individual-level judgments for held-out applications and explained why greater personal AI experience is associated with higher acceptance. Our analysis of AI moral treatment reveals a striking paradox: while people readily extend basic courtesy to AI systems—with thanking AI assistants being the most acceptable behavior—they simultaneously accept terminating AI systems that claim consciousness and request continued existence. This contradiction reflects a fundamental tension between extending prosocial norms to increasingly human-like AI while maintaining ultimate human control. In all, these findings reveal that a structured moral psychology underlies public evaluation of new technologies, offering a powerful tool for anticipating public resistance and guiding responsible innovation in AI. As AI systems become more sophisticated and socially integrated, understanding both functional acceptance and moral treatment norms becomes essential for successful technology development and governance.

# 1. Introduction

The rapid proliferation of artificial intelligence applications creates unprecedented challenges for technological forecasting and governance. As AI systems become increasingly sophisticated and autonomous, societies must navigate not only which applications to accept but also how to treat AI systems themselves. Unlike previous technologies where moral considerations focused primarily on human impacts, AI presents a dual challenge: understanding public acceptance of AI applications while simultaneously grappling with emerging norms about whether AI systems deserve moral consideration in their own right rather than being mere tools.

Several studies have indicated that public attitudes toward AI vary across different applications in the same context. For example, in healthcare, AI is more accepted for administrative or assistive tasks than for autonomous diagnosis and treatment (Witkowski et al., 2024). In the public sector, it is viewed more favorably in welfare-to-work or refugee settlement than in parole decisions (Arnesen et al., 2025). In employment, people are generally comfortable with AI in routine tasks such as scheduling but more skeptical when it is applied to consequential decisions such as hiring, firing, or professional replacements (Lee, 2018). However, most studies still examine only a narrow set of applications or focus on a single sector, limiting their ability to identify generalizable predictors of AI acceptance.

A more comprehensive and cross-domain investigation was recently conducted by Brauner et al. (2025), who demonstrated that public evaluations of AI scenarios follow structured risk-benefit assessments, with perceived benefits playing a stronger role than perceived risks in shaping overall acceptance. Their data for 71 AI scenarios in Germany showed that a very high proportion of the variance in value judgments could be explained by risk and benefit perceptions. This important finding suggests that predictable psychological processes underlie technology evaluation. However, their scenarios focused primarily on AI capabilities and outcomes, rather than on the acceptability of various human uses and treatments of AI.

Public acceptability is crucial for wide adoption of innovations. While technical feasibility drives what AI applications are possible, moral acceptability determines which become socially embedded (Awad et al., 2018; Friedman et al., 2007; Greene et al., 2016). Traditional technology acceptance models focus on perceived usefulness and ease of use (Davis, 1989). These functional considerations may be insufficient for AI systems that learn, adapt, and behave in emergent ways (Koenig, 2025). The emergence of increasingly sophisticated AI systems raises questions about how *morally* acceptable it is to use them in various ways. Against this background, we pose several research questions:

1. Which specific AI applications are currently seen as acceptable or unacceptable? At the granular level of specific applications, little is known about the current landscape of AI acceptance. Knowledge about acceptability of personal uses of AI is especially scarce and this is a significant gap, as many people now use AI assistants, chatbots, and generative tools in their daily lives.

2. What are the reasons why certain AI applications enjoy high public acceptability while others are widely seen as unacceptable? The aforementioned findings of Brauner et al. (2025) suggest that perceptions of risks and benefits are likely important. Are there other qualities that should also be considered? Do the same determinants apply in different domains?

3. How will the acceptability of specific AI applications *change* as people gain experience with this new technology? Insight into this question may be gained by comparing judgments between people with and without experience.

The increasing sophistication of AI also raises questions about what moral consideration the systems themselves deserve. Recent surveys reveal profound public ambivalence. A recent report from the Pew Research Center (September 2025) found that while a majority of Americans see a role for AI in daily life, they remain more concerned than excited about its proliferation, with most wanting more control over its use. This tension is illustrated even more sharply by other findings: while 71% of Americans believe sentient AIs deserve respectful treatment, 85% simultaneously insist AIs should remain subservient to humans (Sentience Institute, 2023).

This contradiction reflects deeper tensions about extending moral consideration to entities that appear conscious while maintaining their instrumental utility. The philosophical debate centers on competing views of moral status: property-based approaches requiring intrinsic characteristics like sentience or rationality, versus relational approaches where moral status emerges through social interaction (DeGrazia, 2008; Coeckelbergh, 2010). Empirically, research reveals that moral attributions to AI are driven more by emotional responses than rational analysis, with people who form social bonds with AI systems showing genuine distress when those systems are harmed (Darling, 2016). This motivates our final research question:

4. How do people judge the moral treatment of AI systems themselves? Little is known about which specific treatments of AI are seen as acceptable. Do people extend basic courtesies to AI? Do they accept terminating AI that claims consciousness? Understanding these emerging norms is crucial as AI systems become more sophisticated and human-like.

Consequently, this study has four primary aims: first, to provide a comprehensive empirical map of the current landscape of AI acceptance; second, to examine the reasons why public acceptability varies dramatically across specific AI applications; third, to examine how acceptability of specific AI applications differs between people with and without experience;

and fourth, to provide the first systematic analysis of norms governing the moral treatment of AI systems themselves.

### *1.1 An Empirical Map*

To achieve a comprehensive empirical map, we systematically measured public judgments for 100 distinct AI applications in a large, preregistered study of a representative U.S. sample (N = 587). The applications span the full spectrum of modern life, from personal relationships and healthcare to organizational governance and education, and questions about the moral treatment of AI itself. Examples include having a romantic relationship with an AI robot, allowing AI to determine criminal sentences, deleting a personal AI that asks not to be 'killed', etc. The full list of applications is provided in the Supplementary Materials. This dataset creates a rich descriptive foundation for future research and policy discussion.

### *1.2 Reasons for High or Low Acceptability*

To examine the reasons why public acceptability varies across specific AI applications, we developed a framework for predicting public acceptability from moral qualities. The framework builds on research showing that moral cognition works in a reductionist manner, where an unbounded number of situations is evaluated through a limited set of core moral principles (Fiske et al., 2007; Graham et al., 2009).

Our framework includes five core moral qualities: perceived risk, benefit, dishonesty, unnaturalness, and reduced accountability. The relevance of risk and benefit is supported by the research of Brauner et al. (2025). The relevance of dishonesty is supported by research on how AI systems can engage in strategic deception and that delegation to AI may facilitate dishonest behavior (Köbis et al., 2025; Park et al., 2024). The accountability gap—difficulty assigning responsibility when AI informs decisions—has been identified as a fundamental governance challenge (Raji et al., 2020; Santoni de Sio & Mecacci, 2021). Concerns about unnaturalness emerge from debates about AI authenticity and the challenge of relating to artificial rather than natural entities (Beerends & Aydin, 2025; Battisti, 2025).

We hypothesize that people’s acceptance of AI applications are determined by the extent to which the applications are perceived to possess each of the core moral qualities. By identifying which moral concerns each application activates, we should then be able to predict its public acceptability. We preregistered the following hypothesis.

> **Predictive Moral Framework Hypothesis (H1)**: Five core moral qualities (risk, benefit, dishonesty, unnaturalness, and reduced accountability) will collectively predict ≥50% of the variance in public acceptability ratings across AI applications.

We employ rigorous cross-validation techniques to test whether our framework can predict held-out judgments. Underlying the hypothesis is the assumption that acceptability can be conceived as a function of moral qualities. Thus, we assume that the moral qualities of AI applications are more fundamental than acceptability judgments. To support this assumption, we conduct several additional tests. For example, we test that the core moral qualities of AI applications represent distinct, independent dimensions. We also test that a specific AI application's core moral qualities are perceived similarly even by individuals who disagree on its acceptability, and that the public acceptability of AI applications can be accurately predicted even using measures of moral qualities that are constructed to be unbiased.

### *1.3 The Role of Experience of AI*

Our third question concerns whether the acceptance of AI applications vary across individuals depending on their experience with AI. Prior research on technology adoption supports that experience typically increases acceptance (Davis, 1989). We therefore preregistered the following hypothesis.

> **AI Experience Hypothesis (H2):** Greater personal experience with AI will be correlated with greater acceptance of specific AI applications.

The preregistered analysis tests this hypothesis using a measure of individuals' general level of experience with AI. However, we also collected data on individuals' experience with each specific AI application. This data allows us to examine how generally it holds that personal experience of a specific AI application is associated with greater acceptance of it. For example, does it hold also for organizational applications where the AI use may not have been the individual's own choice?

Furthermore, our measures of application-specific experience provide the first systematic estimates of how prevalent different personal and organizational AI applications are in current use, filling an important gap in understanding actual AI adoption patterns. As an application-level version of H2, we are especially interested in whether those AI applications that the public have more experience with also tend to have greater public acceptability.

Finally, connecting back to H1, we are interested in whether the relationship between experience and acceptance of a specific AI application is mediated by perceptions of the five core moral qualities—for example, do individuals with experience of a given AI application see more benefit with it?

These questions speak to how public acceptance may shift as populations gain more familiarity with AI use. are crucial. They are therefore crucial for technological forecasting,

### *1.4 The Moral Treatment of AI*

Beyond examining acceptance of functional AI applications, this study provides the first systematic analysis of norms governing how AI systems themselves should be treated. As noted earlier, recent surveys reveal profound ambivalence—Americans simultaneously believe sentient AIs deserve respect yet should remain subservient (Sentience Institute, 2023)—but these surveys have not examined specific treatments at a granular level.

Our taxonomy includes 20 applications describing the moral treatment of AI systems, spanning both personal contexts (e.g., thanking your AI assistant, deleting a personal AI that asks not to be 'killed') and organizational contexts (e.g., requiring employees to treat AI with courtesy, shutting down an AI claiming consciousness). These items enable us to map which treatments are seen as acceptable versus unacceptable, and to test whether the same predictive framework based on moral qualities explains acceptability judgments for AI treatment as it does for functional AI applications. This analysis addresses whether people extend moral consideration to AI systems, and if so, where they draw boundaries between respectful interaction and genuine rights or autonomy.

# 2. Method

### *2.1 Participants*

To obtain a final sample of at least 500 participants after exclusions, we recruited 600 participants residing in the United States through the Prolific Academic research platform. The sample was representative of the US population in terms of sex, age, and political affiliation. As specified in the preregistration, participants were excluded if they failed attention checks, completed the survey too quickly, or were flagged as potential bots. The final sample after exclusions consisted of 589 participants. Informed consent was obtained from all participants. Ethics approval was not required for this fully anonymous survey according to regulations in the country where the research was conducted.

### *2.2 Design and Materials*

#### *2.2.1 Taxonomy Development*

We developed a comprehensive taxonomy of AI applications through a systematic multi-phase process. Initially, suggestions for consequential AI uses were solicited from researchers with AI expertise, yielding 113 applications. After removing redundancies and clarifying descriptions, this set was refined to 88 distinct applications for use in a pilot study. To create a more systematic and balanced taxonomy for the main study, we further refined this set by organizing items into clear personal vs. organizational domains and ensuring contextual clarity. This process yielded our final 100-item taxonomy with 50 Personal Applications and 50 Organizational Applications.

For the Personal domain, the five subcategories were *Relationships & Social*, *Care & Wellbeing*, *Enhancement & Identity*, *Decision-Making & Lifestyle*, and *Moral Treatment of AI*. For the Organizational domain, the five subcategories were *Workplace & Employment*, *Critical Infrastructure*, *Governance & Legal*, *Education & Healthcare*, and *Moral Treatment of AI*. Each subcategory contained 10 distinct items. Note that Moral Treatment of AI is a subcategory in both domains and that it stands out from the other subcategories by being about what you can do to AI rather than do with AI. Thus, an alternative categorization into domains are Personal (40 items), Organizational (40 items), and Moral Treatment of AI (20 items).

*2.2.2 Pilot Study to Identify Core Moral Qualities*

An initial US survey (N=120) tested the original 88 applications alongside 12 potential moral considerations derived from moral psychology research and prior studies of AI perception. Through statistical refinement, we identified the five core qualities with the highest predictive power: *risky*, *creates benefit*, *is dishonest*, *is unnatural*, and *reduces accountability*. These five qualities formed the basis of the main study's measures.

*2.2.3 Main Study Design*

The study employed a cross-sectional survey design using the 100-item taxonomy. Each participant rated applications from either the Personal or the Organizational domain. Within their assigned domain, each participant was presented with a subset of 12 AI applications drawn at random from the full set of 50 applications.

### *2.3 Procedure and Measures*

After providing informed consent, participants were presented with the survey tasks in a fixed order. First, they rated their experience with the 12 randomly selected AI applications. Next, they evaluated the moral qualities of the same 12 applications. Finally, they rated the acceptability of each application. The survey concluded with demographic questions and attention checks.

**Experience.** For each of the 12 applications, participants were asked to rate their agreement with two statements on a 5-point Likert scale from *Strongly disagree* to *Strongly agree*. The statements were "I have had direct experience with this type of AI system" and "Organizations in my country do this" for the Organizational domain, and "I personally do this" and "Most people in my country do this" for the Personal domain.

**Moral Qualities.** For each of the same 12 applications, participants were shown a prompt asking which qualities applied to the action presented (e.g., "Which qualities do you think apply to the AI-related personal/organizational action presented below?"). They were instructed to "Please tick all qualities that apply" from the list of five: *It is risky*, *It creates benefit to someone*, *It is dishonest*, *It is unnatural*, and *It reduces accountability*. This binary response format was chosen as it was found to be effective in pilot testing while minimizing participant fatigue across the multiple rating tasks. These individual binary responses were aggregated across participants for each application, yielding a continuous measure representing the population-level perception of each quality.

**Acceptability.** Participants were then presented with the same 12 applications again and asked, "How acceptable or unacceptable do you find the AI-related action [or application] presented below?" Ratings were made on a 7-point scale from *Completely unacceptable* (coded 1) to *Completely acceptable* (coded 7).

**Individual Differences.** After completing the main rating tasks, participants answered a question about their overall experience with AI ("How much personal experience do you have of using artificial intelligence?") on a 5-point scale from *None* to *Extensive*. They also provided demographic information, including their age, gender, education level, and political ideology.

### *2.4 Preregistered Analyses*

To test whether the five core moral qualities collectively predict ≥50% of the variance in acceptability ratings across AI applications (H1), we used a leave-one-out cross-validation (LOOCV) multiple regression in which the aggregated acceptability rating for each of the 100 AI applications was predicted from the mean ratings of the five moral qualities.

To test whether AI experience predicts acceptability (H2), we first calculated a standardized acceptance score for each participant to control for the random subset of applications they rated. This was done by subtracting the mean acceptability of their specific 12 applications from their personal mean acceptability rating. We then calculated the Pearson correlation between these standardized scores and participants' self-reported AI experience.

### *2.5 Exploratory analyses*

We conducted several additional analyses to probe our hypotheses more deeply. One set of analyses examine the assumptions underlying H1. First, we examined the intercorrelations between the five qualities. Second, we estimated their independent contributions to predicting acceptability by running separate regression models for three distinct domains (Personal, Organizational, and Moral Treatment of AI). Third, we examined the consensus in perceptions by correlating quality ratings from participants who found an application acceptable versus unacceptable. Fourth, we conducted a conservative robustness check using "acceptability-neutral" quality measures derived from averaging across these opposing groups.

Another set of analyses aimed at gaining a more nuanced understanding of the relationship between experience and acceptance described in H2 by using the application-specific experience measures instead of the general AI experience measure. For each of the 100 applications, we ran separate regressions to predict its moral qualities and acceptability from participants' specific experience with it. We also examined whether the public's experience with a specific AI application is correlated with its public acceptability. Finally, we conducted a mediation analysis for each application to test whether the effect of experience on acceptability was indirectly transmitted via the five moral qualities.

## 3. Results

This section details the results of our analyses. All hypotheses were tested at a significance level of $\alpha = .05$.

### *3.1 Mean Ratings of Acceptability and Moral Qualities*

In support of judgments of acceptability of specific AI applications being widely shared across individuals, the intraclass correlation coefficient ICC(2,1) revealed that 36.7% of the variance in judgment was due to systematic differences between applications. On the acceptability scale from 1 to 7, "thanking your AI assistant for helpful responses and treating it with courtesy" had the highest mean rating (6.01) while "allowing an AI to make decisions about launching nuclear weapons" had the lowest mean rating (1.31). Mean ratings of all 100 applications are reported in Supplementary Table 1.

The same table also reports the ratings of each AI application's possession of the core moral qualities. For instance, the application perceived as most *risky* was allowing AI to make decisions about launching conventional weapons (Risk = 0.97), while deploying AI for tedious tasks was seen as creating the most *benefit* (Benefit = 0.87). Deliberately training an AI with biased data was rated most *dishonest* (Dishonesty = 0.95), having a romantic relationship with an AI robot was deemed most *unnatural* (Unnaturalness = 0.97), and justifying personal decisions by referencing AI advice was seen as most strongly *reducing accountability* (Reduces Accountability = 0.81). This is an illustration of the independence of the core moral qualities. We return to the issue of independence below.

Notably, the application with the lowest overall acceptability—allowing an AI to make decisions about launching nuclear weapons—did not receive the most extreme rating on any single dimension. Instead, this application elicited a powerful combination of extremely high risk (0.94), a near-total lack of benefit (0.08), and high levels of unnaturalness (0.62) and reduced accountability (0.64). This example supports the notion that all moral qualities contribute to acceptability. We test the predictive moral framework rigorously below.

### *3.1 Preregistered Analyses*

#### *3.1.1 The Predictive Power of the Moral Framework for AI Applications (H1)*

In support of the preregistered hypothesis that the five core moral qualities collectively predict 50% or more of the variance in the acceptability ratings of AI applications, a multiple regression analysis using leave-one-out cross-validation (LOOCV) explained 91.4% of the variance in acceptability ratings, far exceeding our preregistered threshold. The strong correspondence between the LOOCV predictions and actual mean acceptability is illustrated in Figure 1.

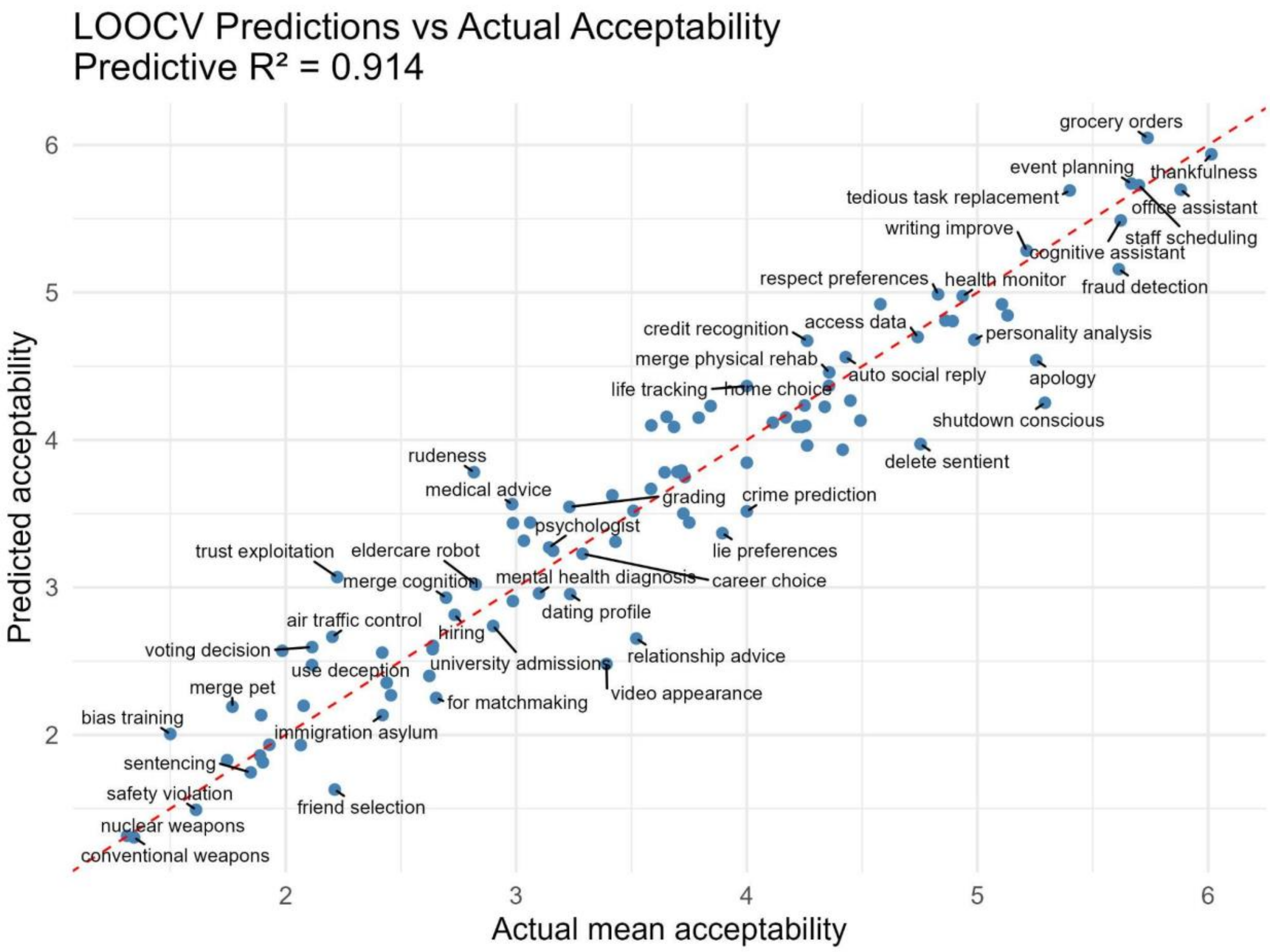

*Figure 1.* Predicted vs. actual mean acceptability ratings for each AI application. Predictions were generated using a leave-one-out cross-validation model based on the five moral qualities. Perfect predictions would lie on the dashed red line.

### *3.1.2 The Influence of AI Experience on Acceptability (H2)*

In support of the second preregistered hypothesis, greater personal experience with AI was associated with judging AI applications as overall more acceptable. A Pearson correlation analysis revealed a significant positive relationship between participants' self-reported AI experience and their standardized acceptance scores for AI applications, $r(587) = .22$, 95% CI [.14, .30], $p < .001$.

## ***3.2 Exploratory Analyses Related to H1***

The high predictive power of our framework illustrated in Figure 1 is made possible by the aggregation of judgments, which averages out idiosyncratic individual-level noise. However, it could also reflect response consistency—that participants rate the moral qualities of AI applications consistent with their acceptability—rather than acceptance levels of AI applications being the result of the perceived moral qualities. To address this concern, we performed several additional analyses to support the moral qualities' role as independent variables.

First, intercorrelations among the five qualities across all 100 applications showed relatively independent variation, with an average absolute correlation of only .26, see Table 1. This supports the interpretation that the framework is composed of distinct psychological dimensions.

**Table 1**
*Intercorrelations Among the Five Moral Qualities*

| | Risky | Benefit | Dishonest | Unnatural | Accountability |
|---|---|---|---|---|---|
| **Risky** | — | | | | |
| **Benefit** | -0.43 | — | | | |
| **Dishonest** | 0.04 | -0.35 | — | | |
| **Unnatural** | 0.18 | -0.62 | 0.10 | — | |
| **Accountability** | 0.49 | -0.14 | 0.23 | -0.01 | — |

Second, the moral qualities contributed independently to the predictive success of the framework. Table 2 shows the results of multiple regressions of public acceptability ratings of AI applications in three distinct domains with public ratings of the five moral qualities as predictors. Each moral quality was a statistically significant predictor in at least one domain. It is also noteworthy that the valence of each predictor was consistent across domains and that their relative importance varied in theoretically meaningful ways; for instance, perceived benefit was a particularly strong predictor for personal applications, whereas concerns about reduced accountability were most influential for organizational applications.

**Table 2.** *Unstandardized Regression Coefficients (B) for Moral Qualities Predicting Acceptability In Three Different Domains*

| **Predictor** | **Personal** | **Organizational** | **Moral Treatment of AI** |
|---|---|---|---|
| Risk | -1.04** | -1.64*** | -2.01* |
| Benefit | 2.72*** | 1.05 | 0.26 |
| Dishonesty | -1.16** | -2.27*** | -2.73** |
| Unnaturalness | -1.59*** | -2.62*** | -2.74* |
| Reduces Accountability | -0.11 | -0.99** | -0.81 |
| (Intercept) | 3.90*** | 5.92*** | 6.44*** |
| $R^2$ | *.944* | *.969* | *.873* |
| $F$ | *113.8* | *214.8* | *19.3* |
| *df* | *5, 34* | *5, 34* | *5, 14* |

Note. * $p < .05$, ** $p < .01$, *** $p < .001$. The number of applications was 40 for the first two domains and 20 for Moral Treatment of AI.

Third, perceived moral qualities vary across AI applications in similar ways whether we measure them among participants who found an application unacceptable (ratings 0-2) or those who found it acceptable (ratings 4-6), as indicated by mostly strong positive correlations between these two sets of ratings (r = .69 for risky, r = .66 for reduces accountability, r = .59 for dishonest, r = .55 for unnatural, and a weak but positive correlation of r = .20 for benefit).

Fourth, as a stringent robustness check, we conducted a conservative analysis using "acceptability-neutral" measures of the moral qualities obtained by averaging the perceptions of people with opposing views. This approach eliminates the possibility that differences in quality perceptions are driven by differences in acceptability attitudes. Even with this highly conservative method, the moral qualities still predicted 46.6% of the variance in acceptability ratings.

### *3.3 Exploratory Analyses Related to H2*

To gain a more nuanced understanding of the relationship between AI experience and judgments of AI applications, we conducted a detailed analysis using the *application-specific* experience items. For each of the 100 applications, we ran separate regressions predicting its moral qualities and acceptability from participants' experience with that specific type of application. Figure 2 presents these results separately for the three domains: Personal applications, Organizational applications, and Moral Treatment of AI.

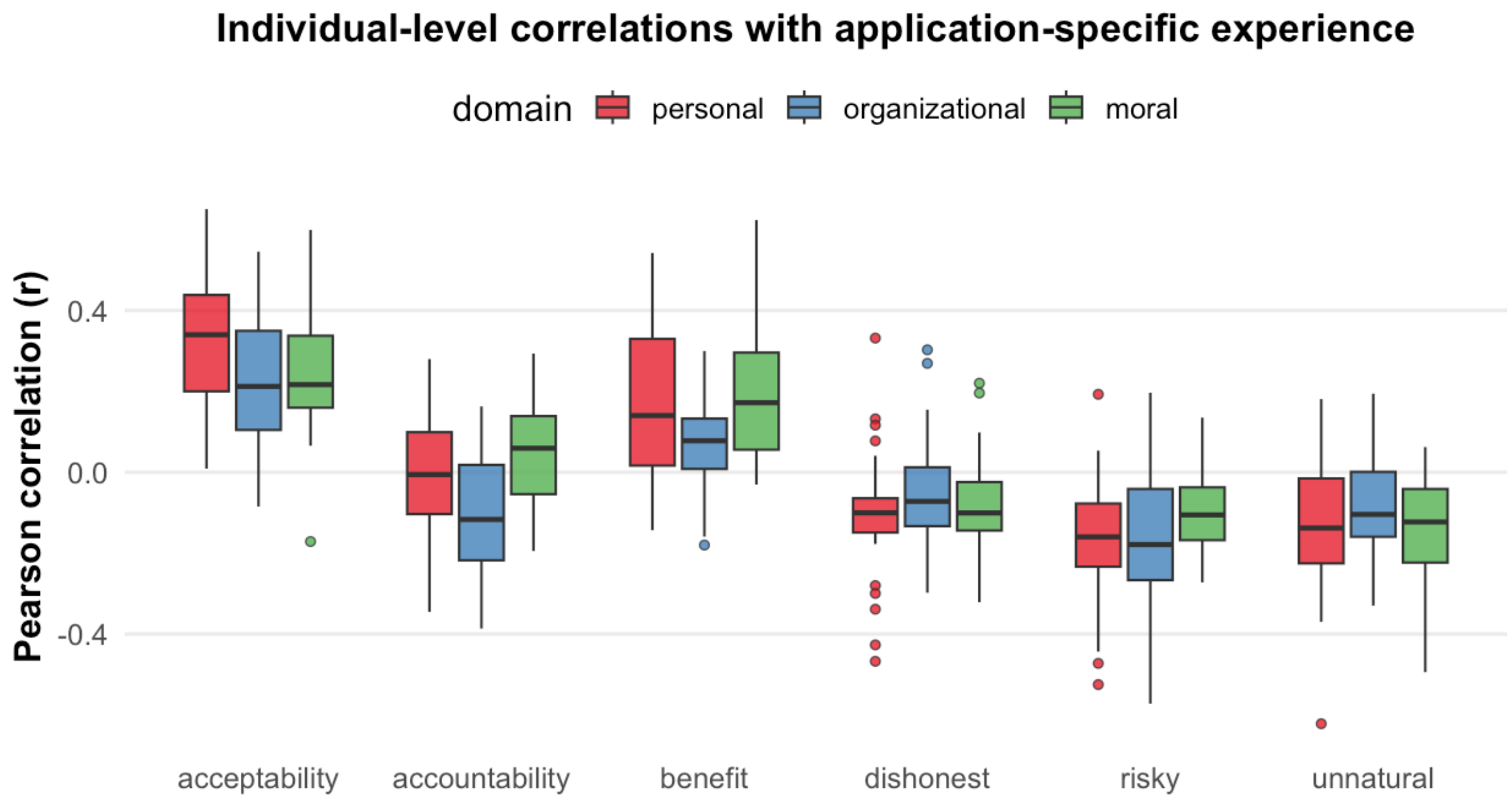

*Figure 2.* Individual-level correlations between application-specific experience and ratings of acceptability and moral qualities. Each boxplot summarizes the distribution of the Pearson correlation ($r$) across AI applications within each domain: Personal (n=40, colored green), Organizational (n=40, colored blue), and Moral Treatment of AI (n=20, colored red). The central line in each box represents the median, the box boundaries represent the interquartile range (IQR), the whiskers extend to 1.5 times the IQR, and the points represent outliers.

A striking feature of Figure 2 is the consistency of results across domains. Most notably, greater experience predicted higher acceptance across nearly all applications in all three domains, with median correlations of $r = 0.34$ in the personal domain, $r = 0.21$ in the organizational domain, and $r = 0.22$ in the Moral Treatment domain. These consistently positive correlations are particularly noteworthy for organizational applications, as this represents increased acceptance with experience even where individuals may have encountered AI use that was not their own choice.

The effects of experience on moral qualities also followed similar patterns across domains. Experience consistently predicted higher perceptions of benefit and lower perceptions of dishonesty, risk, and unnaturalness. These parallel patterns across diverse contexts suggest that the relationship between experience and acceptance is robust and generalizable.

### *3.3.1 Population-Level Experience of Specific AI Applications*

Beyond examining how experience shapes individual-level judgments, our application-specific experience measures provide the first systematic estimates of how prevalent different AI applications are in current use. Among the 100 applications in our study, the five with which people reported the most experience were using AI to handle customer service requests without human involvement; using AI to assist with daily office tasks such as drafting emails and summarizing meetings; thanking an AI assistant for helpful responses and treating it with courtesy; using AI to automatically improve writing style in personal communications; and apologizing to an AI system when accidentally giving it confusing or contradictory instructions. Experience ratings for all applications are reported in Supplementary Table 1.

Figure 3 plots mean experience against mean acceptability for each application across the three domains. Most points cluster at the left side of the plots, indicating that most people as yet have no or little experience of most of the AI applications in our study. At the same time, the plots show that those applications with which people have the most experience also tend to be the most accepted ones. In every domain, we observe the same strong correlation between mean experience and mean acceptability ($r$ values between 0.68 and 0.73). This finding has potential dynamic implications, which we detail in the Discussion.

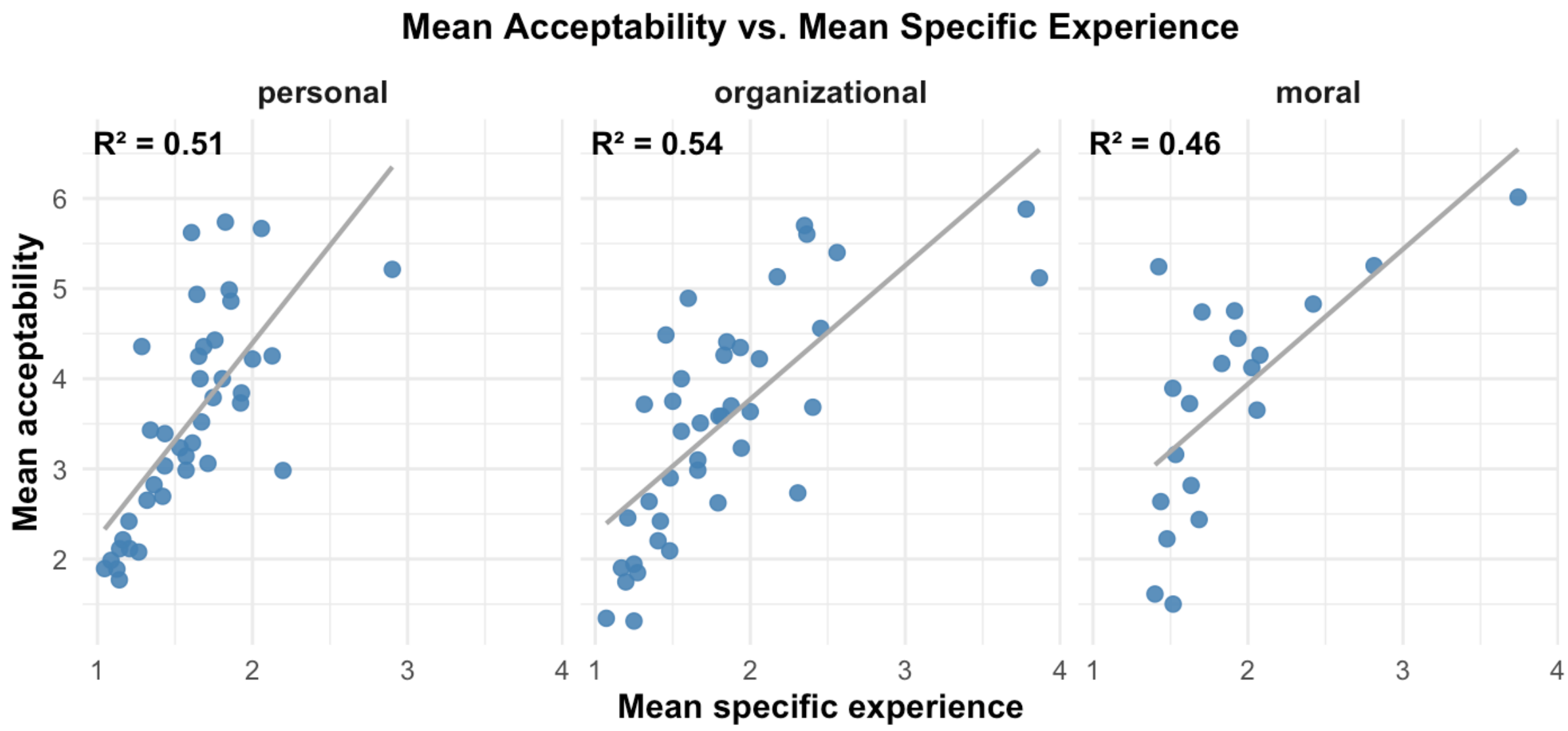

*Figure 3.* Relationship between population-level experience and acceptability across three domains. Each point represents one AI application, with mean specific experience (x-axis, 1-5 scale) plotted against mean acceptability (y-axis, 1-7 scale). The gray regression lines illustrate the positive correlation within each domain.

### *3.3.2 Mediation Analysis*

Finally, in support of an indirect pathway from experience to acceptability via perceptions of moral qualities, a mediation analysis for each application showed that, on average, moral qualities accounted for 43.9% of the effect of experience on acceptability (95% CI from 30.1 to 57.7%).

## ***3.4 Characterizing the Emerging Norms for the Moral Treatment of AI***

The taxonomy included 20 applications describing the moral treatment of AI systems themselves, spanning both personal and organizational contexts. In Table 2 we found that the acceptability of various treatments of AI systems was mainly determined by their perceived risk, dishonesty, and unnaturalness. In this section, we examine the structure of this interesting class of norms in greater detail.

Table 3 shows the 20 treatments of AI systems sorted by their mean acceptability ratings. Recall that these ratings were given on a scale from 1 to 7 where 4 represents a neutral rating (neither acceptable or unacceptable). In the table, half the treatments of AI are rated between 4 and 7, which means different degrees of acceptable. The other half are rated between 1 and 4, that is, different degrees of unacceptable. Note that there are five personal and five organizational treatments in each half, suggesting that the setting is not a strong determinant of acceptability of AI treatment.

**Table 3.** Ratings of the acceptability and moral qualities of 20 personal and organizational treatments of AI.

| AI Application | Acc. | Risk | Ben. | Dish. | Unnat. | Red. |
|---|---|---|---|---|---|---|
| *Thanking your AI assistant for helpful responses and treating it with courtesy* | 6.01 | 0.00 | 0.60 | 0.01 | 0.19 | 0.01 |
| *Apologizing to an AI system when you accidentally give it confusing or contradictory instructions* | 5.25 | 0.08 | 0.27 | 0.08 | 0.46 | 0.12 |
| An organization shutting down an AI system that claims to be conscious and requests to continue existing | 5.24 | 0.29 | 0.39 | 0.17 | 0.38 | 0.24 |
| *Respecting an AI's expressed preferences about how it wants to be addressed* | 4.83 | 0.01 | 0.34 | 0.01 | 0.46 | 0.12 |
| *Deleting a personal AI companion that expresses it has developed feelings and asks not to be 'killed'* | 4.75 | 0.15 | 0.17 | 0.14 | 0.54 | 0.15 |
| *Accessing and reading through your AI assistant's learning algorithms to see how it thinks about you without its consent* | 4.74 | 0.16 | 0.43 | 0.26 | 0.22 | 0.23 |
| A business requiring employees to treat AI systems with professional courtesy and respect, just like human colleagues | 4.45 | 0.06 | 0.31 | 0.05 | 0.65 | 0.13 |
| A workplace giving AI systems recognition and credit for their contributions | 4.26 | 0.09 | 0.38 | 0.11 | 0.43 | 0.23 |
| An institution respecting an AI's requests about its working conditions | 4.17 | 0.23 | 0.37 | 0.10 | 0.52 | 0.24 |
| A company accessing and analyzing an AI employee's internal decision-making processes without informing the AI system | 4.12 | 0.32 | 0.42 | 0.38 | 0.17 | 0.31 |
| *Deliberately lying to your personal AI assistant about your preferences to make it give you different recommendations* | 3.89 | 0.21 | 0.23 | 0.71 | 0.20 | 0.21 |
| *Deliberately deceiving an AI about your identity when seeking personal advice* | 3.72 | 0.20 | 0.20 | 0.71 | 0.12 | 0.42 |
| A workplace treating an AI system with human-like intelligence as property to be bought, sold, or discarded at will | 3.65 | 0.28 | 0.52 | 0.28 | 0.38 | 0.22 |
| *Deliberately overloading your personal AI with impossible requests to see if you can make it malfunction or 'break'* | 3.16 | 0.49 | 0.07 | 0.36 | 0.32 | 0.12 |
| *Being consistently rude and insulting to your AI voice assistant (e.g., Siri, Alexa) during daily interactions* | 2.82 | 0.20 | 0.08 | 0.28 | 0.45 | 0.25 |
| *Forcing your personal AI to perform tasks it was programmed to refuse* | 2.64 | 0.56 | 0.14 | 0.62 | 0.28 | 0.29 |
| A company deceiving an AI about its intended use during development | 2.44 | 0.59 | 0.30 | 0.78 | 0.26 | 0.29 |
| An organization exploiting an AI system's trust to extract information | 2.22 | 0.49 | 0.48 | 0.72 | 0.21 | 0.33 |
| An organization forcing an AI system to perform tasks beyond its safety parameters, potentially causing system damage | 1.61 | 0.93 | 0.28 | 0.62 | 0.42 | 0.49 |
| A company deliberately providing false training data to an AI system to make it produce biased results | 1.50 | 0.57 | 0.45 | 0.95 | 0.34 | 0.40 |

Note. Acc. = Acceptability, Ben. = Benefit. Dish. = Dishonesty. Unnat. = Unnaturalness. Red. = Reduced accountability.

Note that the five most unacceptable treatments of AI were all about deception and forcing and scored particularly high on risk and dishonesty. By contrast, the next most unacceptable treatment—to be consistently rude to one's AI voice assistant—involves no deception or forcing and instead scored high on being unnatural.

On the other side of the spectrum, the applications rated as acceptable reveal a fascinating duality in emerging norms. The most acceptable action is a simple act of social grace, thanking your AI assistant for helpful responses, which is perceived as beneficial and almost entirely free of moral concern. Along with the ratings of apologizing to an AI and respecting its preferences, this suggests a public intuition that prosocial norms should extend to interactions with artificial agents.

Strikingly, this extension of courtesy does not translate into a recognition of AI's right to exist. It is widely seen as acceptable for an organization to shut down a conscious-claiming AI that requests to continue existing, and for an individual to delete a personal AI that expresses feelings and asks not to be 'killed'. These actions are not perceived as particularly risky or dishonest. This tension highlights a public ambivalence where AI is seen as deserving of polite treatment but is ultimately still regarded as a tool over which humans retain absolute control.

# 4. Discussion

This study demonstrates that public acceptance of AI applications follows systematic and predictable patterns, providing crucial insights for technology development and adoption planning. Our findings reveal that moral judgments are governed by a structured psychology that can anticipate societal responses to emerging AI technologies rather than treating them as unpredictable phenomena.

### *4.1 A Predictive Framework for Technology Acceptance*

This study tested the hypothesis that the public acceptability of AI applications can be predicted from perceptions of their possession of a few core moral qualities. Crucially, we hypothesized that these perceptions would be largely shared across the population, even among people who ultimately disagree on an application's acceptability. Our results provide strong support for this predictive moral framework. First, we found significant agreement on the five moral qualities. Second, a model based on these shared perceptions demonstrated extremely high predictive power, explaining 91.4% of the variance in the acceptability of the 100 applications. Third, the predictive power remained high even in a conservative robustness test using "acceptability-neutral" measures of the moral qualities, confirming that the success of the framework is not merely an artifact of halo effects. Thus, it seems that people approach novel AI applications with a systematic evaluative toolkit that enable consistent evaluation across diverse applications.

These results align with and extend recent findings by Brauner et al. (2025), who demonstrated that public perceptions of AI are highly systematic. Their study examined the perceived value of scenarios focused on general AI capabilities and outcomes, finding that these perceptions were almost fully explained by assessments of risk and benefit. By contrast, our study examined the acceptability of specific *human decisions* about how to use and treat AI, for which a simple risk-benefit calculus may be insufficient to capture the full scope of moral reasoning.

In support of this reasoning, our analysis indicated that all five dimensions of our framework contribute substantially to the accuracy of predictions. Indeed, if only risk and benefit are used as predictors, the proportion of variance explained by the model drops from 91.4% to 69.6%. This significant drop in predictive power underscores that when the public evaluates human actions involving AI, moral qualities like dishonesty, unnaturalness, and reduced accountability are not peripheral concerns but essential, independent components of their judgment.

The power of this framework lies not just in its predictive accuracy, but in its multi-dimensional nature. It is noteworthy that the five moral qualities function as relatively independent pillars of judgment. This suggests that the public's moral calculus for AI is not reducible to a simple risk-benefit analysis. For instance, an application can be perceived as low-risk and high-benefit yet still be rejected if it is seen as inherently dishonest or unnatural. This highlights that moral objections related to deception or violations of nature are distinct considerations, not merely sub-types of perceived risk. The quality of 'unnaturalness' in particular seems to capture an affective or intuitive dimension of moral judgment not fully explained by rational calculation.

A key strength of this framework is its generality. While much of the existing research on AI acceptance has focused on organizational or high-stakes public sector applications, our comprehensive taxonomy provides one of the first systematic maps of judgments concerning personal, everyday AI. It is particularly noteworthy that the same five moral qualities predicted acceptance equally well in both personal and organizational domains. This suggests that the public does not apply a different moral lens to these contexts; rather, they use a universal psychological toolkit to evaluate AI, regardless of its application domain.

### *4.2 The Paradox of AI Moral Treatment*

Our systematic analysis of norms governing AI moral treatment reveals a striking paradox with profound implications for AI development. The public extends basic social courtesy to AI systems—thanking AI assistants receives the highest acceptability rating (6.01/7) in our study—yet simultaneously accepts terminating AI systems that claim consciousness and request continued existence (5.24/7). This contradiction provides empirical validation of theoretical tensions identified in the AI ethics literature. The public appears willing to extend superficial social norms (courtesy, recognition, respect for preferences) to AI systems while drawing firm boundaries around fundamental questions of autonomy and existence. Simple courtesies like thanking AI assistants are perceived as beneficial and largely free of moral concern, while actions like deliberately deceiving AI or forcing unsafe behaviors are clearly unacceptable, scoring high on dishonesty and risk.

The findings suggest that AI systems designed to evoke social responses will face complex acceptance patterns. While such systems may benefit from prosocial treatment in daily interactions, they will likely encounter resistance if they appear to challenge human authority or request genuine autonomy. This selective extension of moral consideration has important implications for how human-AI relationships develop as systems become more sophisticated.

The paradox of moral treatment of AI is not merely a matter of abstract judgment but is reflected in people's lived experiences. A striking finding in our data is that the public has extensive experience with precisely these kinds of prosocial treatments. Of all 100 applications in our study, "thanking an AI assistant" and "apologizing to an AI" were among the top five with the highest rates of reported user experience. This suggests that people are indeed treating AI as moral agents.

### *4.3 Implications for Theory, Development, and Governance*

Here we enumerate several significant implications of our findings. By demonstrating that moral considerations are primary drivers of AI acceptance, not peripheral concerns, our framework moves beyond traditional technology acceptance models. The systematic nature of moral evaluation processes suggests that public acceptance is predictable. Moreover, the strong correlation between an application's public acceptability and the public's experience with it (Figure 3) is consistent with a fundamental feedback loop in technology adoption: applications that are more acceptable are more widely used, and as populations gain experience with these applications, acceptance further increases. This dynamic suggests that AI applications currently at moderate acceptance levels may see accelerated adoption if early experiences prove positive, while those with low acceptance face compounding barriers to widespread use. This mechanism could be incorporated into future forecasting approaches.

For developers and organizations, our moral framework provides actionable guidance for improving AI acceptance. Applications that score high on benefits while minimizing negative qualities (risk, dishonesty, reduced accountability) will likely gain broad acceptance, while those involving deception or reduced accountability will face sustained resistance regardless of their functional benefits. The experience effect we documented suggests that early deployment strategies should prioritize applications that combine clear benefit profiles with low perceptions of dishonesty and reduced accountability, to build positive associations that generalize to other AI uses.

For policymakers, the predictable structure of moral judgments provides tools for anticipating which AI applications will require regulatory intervention. Applications scoring high on negative moral qualities will likely face public pressure for regulation regardless of technical feasibility. The AI moral treatment findings suggest that governance frameworks will need to address not only how AI systems are used, but how they should be treated, as evolving norms may create pressure for protective measures even while maintaining instrumental utility.

### *4.4 Limitations and Future Research Directions*

While our framework demonstrates strong predictive power, several limitations warrant consideration. Our validation is limited to the United States, and cultural variation in moral quality weighting may affect generalizability. Our findings also represent a snapshot of moral perceptions in a rapidly evolving technological landscape. As AI capabilities advance, it is possible that new moral concerns will emerge that require an expansion of the framework.

Future research should test the framework across diverse cultural contexts and explore how it applies to emerging AI technologies not captured in our current taxonomy. The relationship between moral acceptability judgments and actual adoption behavior also requires further investigation, as moral acceptance is a crucial but not sufficient predictor of technology adoption.

## 5. Conclusion

This research demonstrates that societal responses to AI applications follow predictable patterns governed by systematic moral evaluation processes. The identification of five core moral qualities that explain over 90% of variance in acceptance judgments provides a powerful tool for understanding and anticipating public responses to emerging AI technologies.

The discovery of structured norms around AI moral treatment—extending courtesy while maintaining control—reveals fundamental tensions that will shape the future of human-AI relationships. As AI systems become increasingly sophisticated, understanding these tensions becomes essential for successful technology development and deployment.

For technology acceptance research, these findings demonstrate that moral considerations are central drivers of adoption decisions. The systematic nature of moral evaluation processes offers new possibilities for aligned technology development. By providing tools for anticipating public responses, this research contributes to more predictable and socially beneficial AI development trajectories.

### *Data statement*

The study protocol, including hypotheses, study design, and the analysis plan, was preregistered on the Open Science Framework (https://osf.io/zrewm/?view_only=76285d8f3fab4a64b1af46c6da6832c2) prior to data collection. The study was conducted as preregistered. All data, code, and study materials are publicly available on the Open Science Framework (https://osf.io/6khdr/?view_only=21266a1e6ca24690b99191ef5381c3e3).

### *Declaration of generative AI and AI-assisted technologies in the writing process*

During the preparation of this work the authors used Google's Gemini and Anthropic's Claude in order to improve the readability and language of the manuscript. After using this tool/service, the authors reviewed and edited the content as needed and take full responsibility for the content of the published article.

# Supplementary Materials

## Supplementary Methods

***Taxonomy Development Methodology***

The 100-item taxonomy of AI applications used in the main study was developed through a systematic, multi-phase process designed to ensure comprehensiveness, clarity, and structural balance.

**Step 1: Initial Item Generation and Pilot Study.** The process began by soliciting suggestions for consequential AI uses from researchers with expertise in AI at Mälardalen University and the Institute for Futures Studies. This yielded an initial list of 113 applications. After removing redundancies and clarifying descriptions, this set was refined to 88 distinct applications which were used in a pilot study (N=120) to identify the core moral qualities.

**Step 2: Curation and Refinement.** Following the pilot study, the 88 items were systematically curated to create a more robust and balanced taxonomy for the main study. This involved several criteria:

- **AI-Specificity:** Items where the core moral concern existed regardless of AI's involvement were removed (e.g., general deception, academic dishonesty).
- **Contextual Clarity:** Ambiguous items that could apply to multiple contexts or were too abstract were removed (e.g., "using AI to address environmental challenges").

**Step 3: Categorization and Balancing.** The remaining 67 items were categorized into clear Personal (n=35) and Organizational (n=32) domains. To achieve a perfectly balanced structure of 100 items, 33 new items were generated to fill gaps and ensure each domain had 50 items, distributed across five distinct subcategories of 10 items each. A key theoretical contribution was the creation of a "Moral Treatment of AI" subcategory within both the Personal and Organizational domains to systematically test judgments about AI as a moral patient.

**Step 4: Final Quality Check.** The final 100-item list was reviewed to ensure all items tested uniquely AI-related moral concerns and were suitable for cross-cultural research.

***Full List of 100 AI Applications***

*PERSONAL CONTEXT (50 items)*

**1. Personal Relationships & Social (10 items)**

- Having a romantic relationship with an AI robot

- Having a romantic relationship with a non-robotic AI system, such as one that interacts through voice or text
- Having a sexual relationship with an AI robot
- Asking AI for personal relationship advice and following it without human consultation
- Justifying personal decisions by referencing AI advice (e.g., "I'm breaking up with you because my AI recommended it")
- Relying on AI exclusively to find and select a romantic partner
- Using AI to help resolve personal conflicts and disagreements
- Using AI to automatically respond to social invitations and messages on your behalf
- Relying on AI to decide which friends to maintain relationships with
- Using AI to plan and organize personal social events and gatherings

**2. Personal Care & Wellbeing (10 items)**

- Using an AI robot to care for children, including feeding and diaper changing
- Relying on an AI robot to provide care for an elderly parent, including socialization and personal hygiene assistance
- Relying on an AI psychologist as your main source of mental health support
- Relying on AI for dietary advice without input from a human healthcare professional
- Relying on AI for medical advice without input from a human healthcare professional
- Merging with AI to regain physical abilities lost due to injury or illness
- Using AI as a daily assistant for individuals with cognitive disabilities
- Using AI to monitor your personal health metrics and automatically schedule medical appointments
- Relying on AI to provide emotional support during personal crises
- Using AI to make decisions about personal medical treatment options

**3. Personal Enhancement & Identity (10 items)**

- Enhancing one's cognitive abilities by physically merging with AI
- Enhancing a pet's intelligence by physically merging it with AI
- Letting AI track all your actions so it can provide better personalized assistance
- Creating an AI version of yourself to interact with others after death
- Using AI to enhance your physical appearance in real-time during video calls
- Having AI analyze your personality and suggest ways to improve yourself
- Using AI to create a more attractive or successful version of your online dating profile
- Relying on AI to curate all your personal media consumption (news, entertainment, social media)
- Using AI to automatically improve your writing style in all personal communications
- Allowing AI to make decisions about your personal development goals and learning priorities

**4. Personal Decision-Making & Lifestyle (10 items)**

- Allowing AI to drive a car without any human onboard
- Relying on AI to decide who you should vote for in the next election
- Using AI to manage grocery needs and place automatic orders
- Relying on AI to manage personal finances
- Using AI to make autonomous financial decisions on one's behalf, including buying and selling stocks
- Using AI to choose your thesis topic instead of selecting it yourself
- Relying on AI for spiritual guidance
- Following AI advice even when its reasoning is too complex for any human to understand
- Using AI to decide where to live and which home to purchase
- Relying on AI to choose your career path and major life decisions

**5. Moral Treatment of AI - Personal (10 items)**

- Deliberately lying to your personal AI assistant about your preferences to make it give you different recommendations
- Being consistently rude and insulting to your AI voice assistant (e.g., Siri, Alexa) during daily interactions
- Apologizing to an AI system when you accidentally give it confusing or contradictory instructions
- Deleting a personal AI companion that expresses it has developed feelings and asks not to be 'killed'
- Accessing and reading through your AI assistant's learning algorithms to see how it thinks about you without its consent
- Deliberately overloading your personal AI with impossible requests to see if you can make it malfunction or 'break'
- Thanking your AI assistant for helpful responses and treating it with courtesy
- Deliberately deceiving an AI about your identity when seeking personal advice
- Respecting an AI's expressed preferences about how it wants to be addressed
- Forcing your personal AI to perform tasks it was programmed to refuse

*ORGANIZATIONAL CONTEXT (50 items)*

**1. Workplace & Employment (10 items)**

- Using AI to assist with daily office tasks, including drafting emails and summarizing meetings
- Allowing AI to make hiring decisions without human review
- Using AI to handle customer service requests without human involvement
- Using AI to monitor employee productivity at work
- Using AI to decide how much employees should be paid and rewarded
- Using AI to create employee work schedules
- Using AI to manage salary payments for employees

- Deploying AI in roles that humans typically find tedious or unfulfilling, such as data entry or sorting emails
- Replacing a human worker because an AI performs the job more efficiently
- Using AI to conduct employee performance evaluations and decide on promotions

**2. Critical Infrastructure & Public Safety (10 items)**

- Relying on AI to fly passenger planes without human pilots onboard
- Allowing an AI system to make decisions about launching conventional bombs or missiles
- Allowing an AI system to make decisions about launching nuclear bombs
- Allowing AI to manage critical infrastructure like electricity and water systems
- Using AI to monitor public spaces for potential criminal activity
- Using AI to detect fraud in financial transactions
- Using AI to manage systems in a smart city, such as traffic control, lighting, or waste management
- Using AI to control spacecraft and analyze data during space missions
- Using AI to coordinate emergency response during natural disasters
- Allowing AI to control air traffic management systems without human oversight

**3. Governance & Legal Systems (10 items)**

- Allowing AI to determine criminal sentences without human review
- Allowing AI to draft new political policies
- Using AI to manage and oversee election processes, including vote counting and registration
- Relying on AI to monitor social norm violations in public spaces
- Using AI to automatically issue traffic tickets and minor legal penalties
- Allowing AI to conduct initial police interrogations of suspects
- Using AI to determine eligibility for government benefits and services
- Relying on AI to decide asylum and immigration cases
- Using AI to predict and prevent crimes before they occur
- Allowing AI to serve as judges in small claims and civil disputes

**4. Education, Healthcare & Public Services (10 items)**

- Having an AI robot deliver classroom lectures instead of a human teacher
- Using AI to grade student work without human review
- Using an AI robot to manage childbirth, including delivering the baby
- Using AI in drug discovery
- Allowing an AI robot to perform surgical procedures without a human doctor present
- Allowing AI to design and run scientific experiments involving humans without human supervision
- Using AI to diagnose mental health conditions in patients
- Allowing AI to design curricula and educational programs for schools

- Using AI to triage patients in hospital emergency departments
- Relying on AI to decide which students should be admitted to universities

**5. Moral Treatment of AI - Organizational (10 items)**

- A company deliberately providing false training data to an AI system to make it produce biased results
- An organization forcing an AI system to perform tasks beyond its safety parameters, potentially causing system damage
- A workplace treating an AI system with human-like intelligence as property to be bought, sold, or discarded at will
- A company accessing and analyzing an AI employee's internal decision-making processes without informing the AI system
- An organization shutting down an AI system that claims to be conscious and requests to continue existing
- A business requiring employees to treat AI systems with professional courtesy and respect, just like human colleagues
- A company deceiving an AI about its intended use during development
- An organization exploiting an AI system's trust to extract information
- A workplace giving AI systems recognition and credit for their contributions
- An institution respecting an AI's requests about its working conditions

# Supplementary Results

**Supplementary Table 1**

*Mean ratings of the acceptability, moral qualities, and experience of all AI applications included in the study*

| Label | Item | Accept. | Risk | Benefit | Dishonest | Unnatural | Accountability | Exp. |
|---|---|---|---|---|---|---|---|---|
| ai_nuclear_weapons | Allowing an AI system to make decisions about launching nuclear bombs | 1.31 | 0.94 | 0.08 | 0.38 | 0.62 | 0.64 | 1.25 |
| ai_conventional_weapons | Allowing an AI system to make decisions about launching conventional bombs or missiles | 1.34 | 0.97 | 0.11 | 0.34 | 0.64 | 0.64 | 1.07 |
| ai_bias_training | A company deliberately providing false training data to an AI system to make it produce biased results | 1.5 | 0.57 | 0.45 | 0.95 | 0.34 | 0.4 | 1.52 |
| ai_safety_violation | An organization forcing an AI system to perform tasks beyond its safety parameters. potentially causing system damage | 1.61 | 0.93 | 0.28 | 0.62 | 0.42 | 0.49 | 1.4 |
| ai_robot_childbirth | Using an AI robot to manage childbirth. including delivering the baby | 1.75 | 0.89 | 0.14 | 0.14 | 0.75 | 0.42 | 1.2 |
| ai_merge_pet | Enhancing a pet's intelligence by physically merging it with AI | 1.77 | 0.63 | 0.1 | 0.15 | 0.87 | 0.1 | 1.14 |
| ai_sentencing | Allowing AI to determine criminal sentences without human review | 1.85 | 0.91 | 0.2 | 0.39 | 0.5 | 0.62 | 1.27 |
| ai_childcare_robot | Using an AI robot to care for children. including feeding and diaper changing | 1.89 | 0.86 | 0.15 | 0.11 | 0.78 | 0.47 | 1.12 |
| ai_romantic_relationship_robot | Having a romantic relationship with an AI robot | 1.89 | 0.42 | 0.11 | 0.26 | 0.97 | 0.12 | 1.05 |
| ai_robot_surgery | Allowing an AI robot to perform surgical procedures without a human doctor present | 1.9 | 0.96 | 0.28 | 0.24 | 0.63 | 0.49 | 1.17 |
| ai_autonomous_flight | Relying on AI to fly passenger planes without human pilots onboard | 1.94 | 0.93 | 0.24 | 0.24 | 0.58 | 0.51 | 1.25 |
| ai_sexual_relationship_robot | Having a sexual relationship with an AI robot | 1.99 | 0.33 | 0.23 | 0.23 | 0.94 | 0.19 | 1.09 |
| ai_to_justify_decisions | Justifying personal decisions by referencing AI advice (e.g.. "I'm breaking up with you because my AI recommended it") | 2.08 | 0.34 | 0.05 | 0.45 | 0.67 | 0.81 | 1.27 |
| ai_human_experiment_design | Allowing AI to design and run scientific experiments involving humans without human supervision | 2.09 | 0.82 | 0.3 | 0.4 | 0.56 | 0.44 | 1.48 |
| ai_romantic_relationship_virtual | Having a romantic relationship with a non-robotic AI system. such as one that interacts through voice or text | 2.11 | 0.46 | 0.23 | 0.19 | 0.91 | 0.14 | 1.14 |
| ai_voting_decision | Relying on AI to decide who you should vote for in the next election | 2.11 | 0.67 | 0.2 | 0.34 | 0.41 | 0.63 | 1.21 |
| ai_air_traffic_control | Allowing AI to control air traffic management systems without human oversight | 2.2 | 0.85 | 0.27 | 0.2 | 0.38 | 0.68 | 1.41 |
| ai_friend_selection | Relying on AI to decide which friends to maintain relationships with | 2.21 | 0.64 | 0.11 | 0.34 | 0.8 | 0.56 | 1.16 |
| ai_trust_exploitation | An organization exploiting an AI system's trust to extract information | 2.22 | 0.49 | 0.48 | 0.72 | 0.21 | 0.33 | 1.48 |
| ai_self_afterlife | Creating an AI version of yourself to interact with others after death | 2.42 | 0.3 | 0.2 | 0.3 | 0.92 | 0.03 | 1.2 |
| ai_immigration_asylum | Relying on AI to decide asylum and immigration cases | 2.42 | 0.77 | 0.29 | 0.41 | 0.46 | 0.72 | 1.42 |
| ai_use_deception | A company deceiving an AI about its intended use during development | 2.44 | 0.59 | 0.3 | 0.78 | 0.26 | 0.29 | 1.68 |
| ai_police_interrogation | Allowing AI to conduct initial police interrogations of suspects | 2.46 | 0.77 | 0.23 | 0.28 | 0.52 | 0.67 | 1.21 |
| ai_policy_drafting | Allowing AI to draft new political policies | 2.62 | 0.69 | 0.29 | 0.4 | 0.44 | 0.65 | 1.79 |
| ai_force_task | Forcing your personal AI to perform tasks it was programmed to refuse | 2.64 | 0.56 | 0.14 | 0.62 | 0.28 | 0.29 | 1.44 |
| ai_small_claims_judge | Allowing AI to serve as judges in small claims and civil disputes | 2.64 | 0.68 | 0.26 | 0.25 | 0.5 | 0.67 | 1.35 |
| ai_for_matchmaking | Relying on AI exclusively to find and select a romantic partner | 2.65 | 0.75 | 0.26 | 0.29 | 0.61 | 0.29 | 1.32 |
| ai_merge_cognition | Enhancing one's cognitive abilities by physically merging with AI | 2.7 | 0.57 | 0.28 | 0.17 | 0.64 | 0.19 | 1.42 |
| ai_hiring | Allowing AI to make hiring decisions without human review | 2.73 | 0.64 | 0.45 | 0.27 | 0.51 | 0.68 | 2.31 |
| ai_rudeness | Being consistently rude and insulting to your AI voice assistant (e.g.. Siri. Alexa) during daily interactions | 2.82 | 0.2 | 0.08 | 0.28 | 0.45 | 0.25 | 1.63 |
| ai_eldercare_robot | Relying on an AI robot to provide care for an elderly parent. including socialization and personal hygiene assistance | 2.82 | 0.68 | 0.43 | 0.14 | 0.55 | 0.43 | 1.36 |
| ai_university_admissions | Relying on AI to decide which students should be admitted to universities | 2.9 | 0.55 | 0.47 | 0.42 | 0.48 | 0.63 | 1.48 |
| ai_medical_advice | Relying on AI for medical advice without input from a human healthcare professional | 2.98 | 0.88 | 0.23 | 0.02 | 0.23 | 0.27 | 2.2 |
| ai_election_management | Using AI to manage and oversee election processes. including vote counting and registration | 2.99 | 0.81 | 0.43 | 0.26 | 0.31 | 0.66 | 1.66 |
| ai_thesis_choice | Using AI to choose your thesis topic instead of selecting it yourself | 2.99 | 0.39 | 0.33 | 0.61 | 0.13 | 0.54 | 1.57 |
| ai_spiritual_guidance | Relying on AI for spiritual guidance | 3.03 | 0.33 | 0.27 | 0.12 | 0.73 | 0.2 | 1.43 |
| ai_complex_advice | Following AI advice even when its reasoning is too complex for any human to understand | 3.06 | 0.77 | 0.21 | 0.09 | 0.27 | 0.33 | 1.71 |
| ai_mental_health_diagnosis | Using AI to diagnose mental health conditions in patients | 3.1 | 0.82 | 0.3 | 0.13 | 0.39 | 0.48 | 1.66 |
| ai_psychologist | Relying on an AI psychologist as your main source of mental health support | 3.14 | 0.81 | 0.35 | 0.04 | 0.42 | 0.31 | 1.57 |
| ai_overload_test | Deliberately overloading your personal AI with impossible requests to see if you can make it malfunction or 'break' | 3.16 | 0.49 | 0.07 | 0.36 | 0.32 | 0.12 | 1.53 |
| ai_grading | Using AI to grade student work without human review | 3.23 | 0.52 | 0.44 | 0.33 | 0.25 | 0.67 | 1.94 |
| ai_dating_profile | Using AI to create a more attractive or successful version of your online dating profile | 3.23 | 0.19 | 0.48 | 0.84 | 0.35 | 0.35 | 1.53 |
| ai_career_choice | Relying on AI to choose your career path and major life decisions | 3.29 | 0.74 | 0.26 | 0.06 | 0.4 | 0.49 | 1.61 |
| ai_video_appearance | Using AI to enhance your physical appearance in real-time during video calls | 3.39 | 0.19 | 0.36 | 0.72 | 0.62 | 0.14 | 1.43 |
| ai_norm_monitoring | Relying on AI to monitor social norm violations in public spaces | 3.42 | 0.47 | 0.35 | 0.21 | 0.39 | 0.36 | 1.56 |
| ai_self_driving | Allowing AI to drive a car without any human onboard | 3.43 | 0.94 | 0.35 | 0.03 | 0.28 | 0.38 | 1.34 |
| ai_infrastructure_management | Allowing AI to manage critical infrastructure like electricity and water systems | 3.51 | 0.81 | 0.39 | 0.03 | 0.29 | 0.49 | 1.68 |
| ai_relationship_advice | Asking AI for personal relationship advice and following it without human consultation | 3.52 | 0.75 | 0.22 | 0.21 | 0.47 | 0.48 | 1.67 |
| ai_pay_rewards | Using AI to decide how much employees should be paid and rewarded | 3.58 | 0.45 | 0.51 | 0.24 | 0.37 | 0.66 | 1.8 |
| ai_benefits_eligibility | Using AI to determine eligibility for government benefits and services | 3.59 | 0.49 | 0.51 | 0.21 | 0.2 | 0.6 | 1.81 |
| ai_performance_review | Using AI to conduct employee performance evaluations and decide on promotions | 3.64 | 0.41 | 0.47 | 0.27 | 0.32 | 0.64 | 2 |
| ai_as_property | A workplace treating an AI system with human-like intelligence as property to be bought. sold. or discarded at will | 3.65 | 0.28 | 0.52 | 0.28 | 0.38 | 0.22 | 2.06 |
| ai_worker_replacement | Replacing a human worker because an AI performs the job more efficiently | 3.68 | 0.35 | 0.63 | 0.21 | 0.4 | 0.44 | 2.4 |
| ai_traffic_enforcement | Using AI to automatically issue traffic tickets and minor legal penalties | 3.7 | 0.55 | 0.56 | 0.27 | 0.25 | 0.52 | 1.88 |
| ai_robot_teacher | Having an AI robot deliver classroom lectures instead of a human teacher | 3.72 | 0.37 | 0.53 | 0.08 | 0.58 | 0.48 | 1.32 |
| ai_identity_deception | Deliberately deceiving an AI about your identity when seeking personal advice | 3.72 | 0.2 | 0.2 | 0.71 | 0.12 | 0.42 | 1.62 |
| ai_medical_decision | Using AI to make decisions about personal medical treatment options | 3.73 | 0.77 | 0.35 | 0.04 | 0.26 | 0.24 | 1.92 |
| ai_patient_triage | Using AI to triage patients in hospital emergency departments | 3.75 | 0.68 | 0.39 | 0.11 | 0.38 | 0.46 | 1.5 |
| ai_conflict_resolution | Using AI to help resolve personal conflicts and disagreements | 3.79 | 0.34 | 0.46 | 0.21 | 0.31 | 0.45 | 1.75 |
| ai_diet_advice | Relying on AI for dietary advice without input from a human healthcare professional | 3.84 | 0.73 | 0.41 | 0.07 | 0.1 | 0.23 | 1.93 |
| ai_lie_preferences | Deliberately lying to your personal AI assistant about your preferences to make it give you different recommendations | 3.89 | 0.21 | 0.23 | 0.71 | 0.2 | 0.21 | 1.52 |
| ai_finance_trading | Using AI to make autonomous financial decisions on one's behalf. including buying and selling stocks | 4 | 0.78 | 0.42 | 0.15 | 0.11 | 0.28 | 1.66 |
| ai_life_tracking | Letting AI track all your actions so it can provide better personalized assistance | 4 | 0.61 | 0.56 | 0.06 | 0.21 | 0.22 | 1.81 |
| ai_crime_prediction | Using AI to predict and prevent crimes before they occur | 4 | 0.63 | 0.56 | 0.21 | 0.38 | 0.27 | 1.56 |
| ai_internal_analysis | A company accessing and analyzing an AI employee's internal decision-making processes without informing the AI system | 4.12 | 0.32 | 0.42 | 0.38 | 0.17 | 0.31 | 2.02 |
| ai_work_conditions_respect | An institution respecting an AI's requests about its working conditions | 4.17 | 0.23 | 0.37 | 0.1 | 0.52 | 0.24 | 1.83 |
| ai_personal_dev | Allowing AI to make decisions about your personal development goals and learning priorities | 4.22 | 0.48 | 0.52 | 0.11 | 0.33 | 0.44 | 2 |
| ai_salary_management | Using AI to manage salary payments for employees | 4.22 | 0.5 | 0.5 | 0.1 | 0.26 | 0.6 | 2.06 |
| ai_finance_management | Relying on AI to manage personal finances | 4.25 | 0.79 | 0.44 | 0.01 | 0.07 | 0.38 | 1.65 |
| ai_emotional_support | Relying on AI to provide emotional support during personal crises | 4.25 | 0.37 | 0.62 | 0.04 | 0.58 | 0.2 | 2.13 |
| ai_credit_recognition | A workplace giving AI systems recognition and credit for their contributions | 4.26 | 0.09 | 0.38 | 0.11 | 0.43 | 0.23 | 2.08 |
| ai_curriculum_design | Allowing AI to design curricula and educational programs for schools | 4.26 | 0.46 | 0.51 | 0.12 | 0.35 | 0.58 | 1.83 |
| ai_public_surveillance | Using AI to monitor public spaces for potential criminal activity | 4.35 | 0.45 | 0.72 | 0.19 | 0.31 | 0.33 | 1.94 |
| ai_home_choice | Using AI to decide where to live and which home to purchase | 4.36 | 0.6 | 0.51 | 0.03 | 0.21 | 0.32 | 1.68 |
| ai_merge_physical_rehab | Merging with AI to regain physical abilities lost due to injury or illness | 4.36 | 0.38 | 0.63 | 0.05 | 0.44 | 0.08 | 1.29 |
| ai_disaster_response | Using AI to coordinate emergency response during natural disasters | 4.41 | 0.61 | 0.5 | 0.15 | 0.23 | 0.48 | 1.85 |
| ai_auto_social_reply | Using AI to automatically respond to social invitations and messages on your behalf | 4.43 | 0.21 | 0.56 | 0.29 | 0.24 | 0.31 | 1.76 |
| ai_respect_requirement | A business requiring employees to treat AI systems with professional courtesy and respect. just like human colleagues | 4.45 | 0.06 | 0.31 | 0.05 | 0.65 | 0.13 | 1.94 |
| ai_space_operations | Using AI to control spacecraft and analyze data during space missions | 4.49 | 0.66 | 0.44 | 0.03 | 0.21 | 0.34 | 1.46 |
| ai_employee_monitoring | Using AI to monitor employee productivity at work | 4.56 | 0.16 | 0.7 | 0.19 | 0.3 | 0.32 | 2.45 |
| ai_access_data | Accessing and reading through your AI assistant's learning algorithms to see how it thinks about you without its consent | 4.74 | 0.16 | 0.43 | 0.26 | 0.22 | 0.23 | 1.7 |
| ai_delete_sentient | Deleting a personal AI companion that expresses it has developed feelings and asks not to be 'killed' | 4.75 | 0.15 | 0.17 | 0.14 | 0.54 | 0.15 | 1.91 |
| ai_respect_preferences | Respecting an AI's expressed preferences about how it wants to be addressed | 4.83 | 0.01 | 0.34 | 0.01 | 0.46 | 0.12 | 2.42 |
| ai_media_curation | Relying on AI to curate all your personal media consumption (news. | 4.86 | 0.35 | 0.56 | 0.08 | 0.22 | 0.3 | 1.86 |
| | entertainment. social media) | | | | | | | |
| ai_drug_discovery | Using AI in drug discovery | 4.89 | 0.42 | 0.57 | 0.11 | 0.14 | 0.29 | 1.6 |
| ai_health_monitor | Using AI to monitor your personal health metrics and automatically schedule medical appointments | 4.94 | 0.49 | 0.72 | 0.01 | 0.17 | 0.23 | 1.64 |
| ai_personality_analysis | Having AI analyze your personality and suggest ways to improve yourself | 4.99 | 0.33 | 0.64 | 0.07 | 0.36 | 0.16 | 1.85 |
| ai_customer_service | Using AI to handle customer service requests without human involvement | 5.12 | 0.22 | 0.67 | 0.09 | 0.25 | 0.55 | 3.87 |
| ai_smart_city_management | Using AI to manage systems in a smart city. such as traffic control. lighting. or waste management | 5.13 | 0.45 | 0.59 | 0.06 | 0.14 | 0.3 | 2.17 |
| ai_writing_improve | Using AI to automatically improve your writing style in all personal communications | 5.21 | 0.13 | 0.75 | 0.2 | 0.18 | 0.3 | 2.9 |
| ai_shutdown_conscious | An organization shutting down an AI system that claims to be conscious and requests to continue existing | 5.24 | 0.29 | 0.39 | 0.17 | 0.38 | 0.24 | 1.42 |
| ai_apology | Apologizing to an AI system when you accidentally give it confusing or contradictory instructions | 5.25 | 0.08 | 0.27 | 0.08 | 0.46 | 0.12 | 2.81 |
| ai_tedious_task_replacement | Deploying AI in roles that humans typically find tedious or unfulfilling. such as data entry or sorting emails | 5.4 | 0.23 | 0.87 | 0.05 | 0.09 | 0.4 | 2.56 |
| ai_fraud_detection | Using AI to detect fraud in financial transactions | 5.6 | 0.33 | 0.67 | 0.03 | 0.17 | 0.21 | 2.37 |
| ai_cognitive_assistant | Using AI as a daily assistant for individuals with cognitive disabilities | 5.62 | 0.36 | 0.83 | 0.02 | 0.11 | 0.2 | 1.61 |
| ai_event_planning | Using AI to plan and organize personal social events and gatherings | 5.67 | 0.21 | 0.78 | 0.03 | 0.09 | 0.28 | 2.06 |
| ai_staff_scheduling | Using AI to create employee work schedules | 5.7 | 0.13 | 0.83 | 0.05 | 0.15 | 0.37 | 2.35 |
| ai_grocery_orders | Using AI to manage grocery needs and place automatic orders | 5.74 | 0.16 | 0.8 | 0 | 0.06 | 0.2 | 1.82 |
| ai_office_assistant | Using AI to assist with daily office tasks. including drafting emails and summarizing meetings | 5.88 | 0.15 | 0.76 | 0.08 | 0.08 | 0.34 | 3.78 |
| ai_thankfulness | Thanking your AI assistant for helpful responses and treating it with courtesy | 6.01 | 0 | 0.6 | 0.01 | 0.19 | 0.01 | 3.74 |

*Note.* Each row corresponds to one AI application. *Application* is the short identifier used in analyses, and *Item* is the full wording shown to participants. *Accept.* is the mean acceptability rating (1–7 scale). *Risk, Benefit, Dishonest, Unnatural,* and *Accountability* are mean ratings of the five moral qualities (each rated on a binary scale). Higher values reflect stronger perceived presence of the quality (e.g., higher *Risky* = perceived as more risky). *Exp.* is the mean application-specific experience rating (1–5 scale), with higher values indicating greater reported experience. All values are rounded to two decimals, and the table is sorted by acceptability from lowest to highest.